\documentclass{emulateapj}

\newcommand{\hMpc}{{\ifmmode{h^{-1}{\rm Mpc}}\else{$h^{-1}$Mpc}\fi}}

\newcommand{\hkpc}{{\ifmmode{h^{-1}{\rm kpc}}\else{$h^{-1}$kpc}\fi}}

\newcommand{\hMsun}{{\ifmmode{h^{-1}{\rm {M_{\odot}}}}\else{$h^{-1}{\rm{M_{\odot}}}$}\fi}}

\newcommand{\Msun}{{\ifmmode{{\rm {M_{\odot}}}}\else{${\rm{M_{\odot}}}$}\fi}}

\newcommand{\MNUC}{{\sc muc~}}

\newcommand{\MNUCL} {{{\sc mucl}~}}

\newcommand{\MNUW}{{{\sc muw}~}}

\newcommand{\MNGW}{{{\sc mu2w}~}}

\newcommand{\MNUWH}{{{\sc muwhs}~}}

\newcommand{\MNUALL}{{{\sc muw+mu2w}~}}

\shortauthors{Yepes, Sevilla, Gottl\"ober, Silk}

\shorttitle{Cluster abundance and  WMAP3 normalization}

\begin{document}


\title{Is WMAP3 normalization compatible with the X-Ray cluster abundance?}


\author{ Gustavo Yepes, Raul Sevilla}

\affil{Grupo de Astrof\'\i sica,
Universidad Aut\'onoma de Madrid, Madrid E-28049, Spain}

\email{gustavo.yepes@uam.es, raul.sevilla@uam.es}

\author{Stefan Gottl\"ober}

\affil{Astrophysikalisches Institut Potsdam,
An der Sternwarte 16, 14482 Potsdam, Germany}

\email{sgottloeber@aip.de}

\and

\author{Joseph  Silk}

\affil{Oxford Astrophysics, Denys Wilkinson Building, Keble Road, OX13RH,
  Oxford, United Kingdom}

\email{silk@astro.ox.ac.uk}

\begin{abstract}

  We present the mass and X-ray temperature functions
  derived from a sample of more than 15,000  galaxy clusters of 
 the {\sc MareNostrum Universe}  cosmological SPH simulations.
 In these simulations, we follow 
  structure formation in a cubic volume of $500 h^{-1}$ Mpc on a side
  assuming cosmological parameters consistent with either 
 the first  or third  year WMAP  data and gaussian initial conditions.
  We compare our numerical
  predictions with the most recent   observational estimates of the
  cluster X-ray temperature functions   and find that the low
  normalization cosmological model inferred from  the 3 year WMAP data
  results is barely compatible with  the present  epoch X-ray cluster
  abundances. We can only reconcile the simulations with the 
  observational data if we assume a normalization of the
  Mass-Temperature relation which is  a factor of $\sim 2.5-3 $ smaller  than our
  non-radiative simulations predict. This deviation seems to be too
  large to be accounted by the effects of star formation or cooling in
  the ICM, not taken into account in these simulations.
\end{abstract}




\section{Introduction}



Clusters of galaxies are  strong X-ray emitters   that can be observed
at large distances using  the XMM-Newton and Chandra X-ray telescopes. They
are excellent cosmological probes that  can be used to put strong
constraints on the  matter density of the universe, ($\Omega_M$), the
normalization of primordial density fluctuations ($\sigma_8$) and  the
associated spectral index ($n$). The number of massive clusters in
cold dark matter-dominated cosmologies  is known to be exponentially
dependent on $\sigma_8$ \citep{st02}, as has been extensively confirmed by
simulations.  Therefore, the  determination of  the abundance  of
massive clusters    gives one of the best constraints on the
normalization of the initial   power spectrum of density fluctuations,
provided we adopt gaussian initial conditions. An independent
measurement   of the cosmological parameters comes from the study  of
CMB   anisotropies.  The most recent data from 3 year WMAP satellite
    \citep{wmap3} (WMAP3)  gives a value for $\sigma_8 \sim 0.76\pm 0.05$  to within $1 \sigma$ error.  This is smaller than the previous  value of $\sigma_8= 0.84\pm 0.04$
  estimated from the    first year WMAP data \citep{wmap1} (WMAP1). 
  This difference in the normalization
and the matter content ($\Omega_M=0.24$ vs. $\Omega_M=0.3$),  
translates into large  differences,  up to an order of  magnitude 
 as  we will show in this letter,   in the number density of the  most
  massive  objects formed at present in  the Universe. Other
  recent papers  independently also argue  against  the low 
values of $\sigma_8$  obtained from WMAP3 \citep{evrard, rozo}. 

   In order  to  compare the theoretical cluster  mass function for a
   particular cosmological model with the observed abundance of X-ray
   clusters as a function of the ICM gas temperature, one has to assume
   that  the  Mass-Temperature relation is sufficiently  well  known.
   The main obstacle is the accuracy in the determination  of this
   relation.   Small differences can lead to large  changes  in the
   determination of cluster mass (see e.g \citet{henry:2004}) from the
   X-ray  temperature. Estimations of the M-T relation from gas
   dynamical   simulations show large discrepancies, mainly due to
   numerical  resolution effects as well as to the physics involved (see
 \citet{ascasibar:2006} for a review).  Most previous  numerical studies
on the comparison of  cluster mass functions and X-ray temperature
and/or luminosity functions have either  high numerical resolution and
a low number of objects or larger  statistics but with  very low
resolution.

\begin{table*}[t]
\caption{Main features of the simulations used in this work. $\alpha $ and $\log
M_0$ are the best fit parameters of the  $\log (M_{200}/M_0) =$ $ \alpha
\log (T_X/3\mbox{\rm keV})$ relation.
Errors correspond to $1\sigma$ of the linear fit. 
$\Delta \log M_0$  is the maximum scatter in the normalization
(see text)
 \label{simu}}
\begin{tabular}{lcccccccc}
\hline
\hline
Name & $N_p$ & $\Omega_M$ & $h$ &  $\sigma_8$ & $n$ & $\alpha$ & $ \log M_0$ & $\Delta \log M_0$  \\
\hline
\MNUC & $2\times 1024^3$ & 0.3 & 0.7 &  0.9 &1 &  $1.89 \pm 0.02$ & $14.64\pm 0.01$ & 0.46   \\
\MNUCL &  $2\times 512^3$ & 0.3 & 0.7 &   0.9 &1  & $1.71 \pm 0.02$ & $14.52\pm 0.01$ & 0.35    \\
\MNUWH  & $2\times 512^3$ & 0.24 & 0.73 &   0.8 &0.95 & $1.62\pm 0.05$  & $14.56\pm 0.01$ & 0.31 \\
\MNUW &  $2\times 512^3$ & 0.24 & 0.73  &0.75 & 0.95& $1.65\pm 0.04 $ &  $ 14.54\pm 0.01$ & 0.28 \\
\MNGW & $2\times 512^3$ & 0.24 & 0.73 &  0.75 &0.95 & $1.65\pm0.05$  & $14.56\pm0.01 $ & 0.28  \\
\MNUALL & $2\times 512^3$ & 0.24 & 0.73 & 0.75 &0.95 &  $1.60\pm 0.04$  & $14.54 \pm 0.01$ & 0.28  \\
\hline
\hline

\end{tabular}

\end{table*}

The aim of this letter is to study the X-ray cluster Temperature
Function (XTF)  obtained from a set of large-scale non-radiative gas dynamical simulations with sufficient  numerical resolution  and statistics 
to cover the range of temperatures for which observational estimates
of the cluster abundance are known. Our main goal is to test whether
the observed number of   X-ray emitting galaxy clusters can be obtained in 
a  cosmological model  with parameters consistent with WMAP3 or WMAP1 
data at the present time.   For this purpose, we compute the XTF
directly from simulations   and compare them with the most recent
observational estimates. At the same time, we derive the 
values for  the normalization of the M-T relation that best fit the
simulation mass  functions to the observed XTF and compare them with
the M-T resulting from the simulations.

\section{Simulations}

To study the X-ray cluster abundance, we have performed a series of
non-radiative SPH simulations with the {\sc gadget2} code
\citep{Springel05} at the Barcelona Supercomputer Center. Starting at
redshift $z=40$, we followed the non-linear evolution of structures in
gas and dark matter (DM) to the present epoch ($z=0$)
within a comoving cube of $500\hMpc$ on a side. The so-called {\sc
  MareNostrum universe} is the SPH simulation with $2\times 1024^3$
particles (\MNUC).  We assumed a concordance cosmological model with
the following parameters:  total matter density $\Omega_m=0.3$, the
baryon density $\Omega_b=0.045$, cosmological constant
$\Omega_\Lambda=0.7$, Hubble parameter $h=0.7$,  slope of the
power spectrum $n=1$ and  normalization $\sigma_8=0.9$. We 
also ran the same simulation with exactly the same initial data
but lower mass resolution ($2\times 512^3$, \MNUCL) as described in
\citet{gottloeber:2007}.

After the release of the 3-year WMAP data, we 
complemented our numerical data set with new simulations of the same
computational box but using WMAP3 cosmological parameters:
$\Omega_m=0.24$, $\Omega_b=0.0418$, $\Omega_\Lambda=0.76$, $h=0.73$,
$n=0.95$ and $\sigma_8=0.75$.
We have changed both $\Omega_M$ and $\sigma_8$ (rather than only $\sigma_8$) so as to
remain on the  WMAP degeneracy line for these two parameters.
 As in the concordance model, the power
spectrum was kindly provided by Wayne Hu, who computed it by direct
numerical integration of the Boltzmann code. We generated the initial
conditions for the WMAP3-compatible simulations with 
$2\times 512^3$ (\MNUW)
particles in exactly the same way as for the {\sc MareNostrum
  Universe}.  In order to study the effects of cosmic variance, we have
completed a second simulation with a different random realization
(\MNGW). Furthermore, and driven by the results obtained for the XTF
from these simulations, we have also repeated the {\sc MareNostrum Universe}
realization of the WMAP3 cosmology, but with a higher normalization of
the initial power spectrum ($\sigma_8 = 0.8$), consistent within $1
\sigma$ with the WMAP3 best fit (\MNUWH).  In Table \ref{simu}, we summarize
the main characteristics of the simulations and the corresponding
acronyms for reference in what follows. The best fit values of the
 Mass-Temperature relations from clusters obtained in each simulation 
are also shown in the last two columns (see \S{} \ref{xtfSec}).
The clusters have been identified in the simulations by means of a
hierarchical Friends-of-Friends (FOF) halo finder as described in
\citet{gottloeber:2007}. 
For comparison with
observational data, we have estimated total masses (dark + gas) of
clusters at different spherical overdensities (200, 500, 2500)  with respect
to the critical density. To this end, we started at the position 
of the most massive substructure of the
clusters identified with FOF and used the Bound Density Maxima algorithm
\citep{klypin:1999} to find the spherical overdensities.

\begin{figure}
\plotone{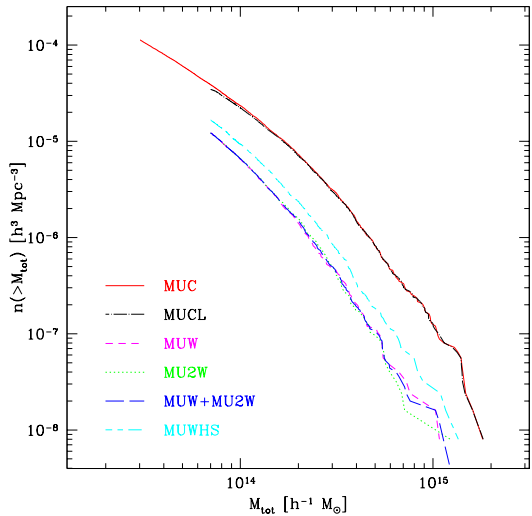}
\caption{Cumulative mass function  from the simulations described in
   the text and Table \ref{simu}
\label{massfunc}}
\end{figure}

\section{Cluster Mass Functions}
In Fig. \ref{massfunc} we plot the resulting cumulative mass functions
for all the simulations described in Table \ref{simu}. In this figure, 
the total mass of objects corresponds to the region enclosing an
overdensity of 200
around the center of mass found as described in the previous section.
As can be deduced  from  this figure, there are no
significant resolution effects on the number of objects as a function of
mass. The mass functions for simulations \MNUC and \MNUCL nicely
overlap each other despite the fact that they differ by a factor of 8
in mass resolution and a factor of $\sim3$ in spatial resolution.  On the
other hand, there is a significant difference in the number of
cluster-size objects depending on the cosmological model.  The number
density of clusters  with masses  $M_{200} \geq 5\times 10^{14} \hMsun$ in
both simulations with the low normalization, best-fit WMAP3 
cosmological parameters, \MNUW and \MNGW,  is $\sim 10$ times  smaller 
 than for the  simulations of the concordance $\Lambda CDM$ model. The \MNUWH
 simulation with  $\sigma_8=0.8$  has a number density  $\sim 2$ higher than
 the simulations with $\sigma_8=0.75$, but still is a factor of $\sim 5$
 smaller than in the concordance cosmology.  
 Finally, Fig \ref{massfunc} also shows that
the effects of cosmic variance is not important in determining the
abundance of clusters at these scales. The agreement of the  mass functions
for the two different realizations of the WMAP3 cosmological model
clearly confirms this. On the other hand, we have also checked for the
possible effects of small volume sampling in the determination of the
mass function for the most massive objects.  To this end, we have
compared our  mass function for the  \MNUWH simulation with the mass
function obtained from a   dark matter only  simulation
 of the same cosmological model as \MNUWH  and number of particles
 but larger computational volume  (1.5 Gpc). This simulation has been
 done also at MareNostrum with the {\sc gadget2} code by P. Fosalba for the
 Dark Energy  Survey project.  The agreement between the two mass
functions is remarkable for halos with masses $M_{200} > 5\times 10^{14} \hMsun$.  

Therefore, we conclude that the estimation of the cluster mass
function from our simulations is robust and not likely to be affected
by numerical effects.  Now, our purpose is to compare them with data
coming from X-ray observations of clusters. As our simulations include
gas dynamics, we can directly measure the X-ray temperature from the
gas content of our halos. In DM-only simulations, one has to rely on
the Mass-X-ray Temperature relation to transform mass into temperature
or vice versa. Here we will do the same exercise and compare the 
calculated XTFs.

\section{X-Ray Temperature Function}
\label{xtfSec}

The most recent published data for the XTF of nearby clusters uses
 temperatures derived from X-ray observations mainly 
 by the  ASCA satellite  \citep{ikebe:2002, henry:2004}
as a measure of the mean temperature of the ICM. 
The differences shown in the temperatures of clusters 
 from these two datasets reflect the systematic errors in the observed XTF. 
For our simulated clusters, we computed several temperature
estimations: the  emission-weighted temperature, $T_ew$ by weighting the
temperature of each SPH particle within
the cluster  by their X-Ray luminosity.
We also computed the spectroscopic temperature,
$T_X$, following the procedure described in \citet{vik:06}, which is
suppose to give a more accurate value of the observed  temperature of an
X-ray emmitting plasma. Therefore, in what follows we will use $T_X$
for the  simulated clusters.

\begin{figure}
\plotone{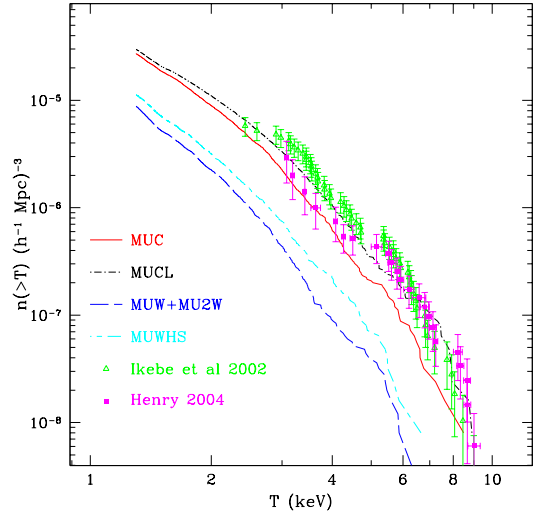}
\caption{Cumulative X-Ray Temperature function for two different cosmologies.
Same notation as in Fig. \ref{massfunc}.
Solid triangles and squares represent \cite{ikebe:2002} and \cite{henry:2004}
data, respectively. Errors were computed as explained in \cite{henry:2004}.
\label{xtf}}
\end{figure}

In Fig. \ref{xtf} we show the cumulative XTF  as a function of the
spectroscopic  $T_X$  for the   clusters found in 
 simulations described in Table \ref{simu}. We also represent the
observational data as points with error bars as described in
\cite{henry:2004}.  The observational data were rescaled to units of
$h=1$.

The predicted number density of X-ray clusters above a given
temperature for the \MNUC  and \MNUCL simulations with  $\sigma_8 =0.9$ is
in good agreement with the data. Again, as in the case of mass, 
the WMAP3 most favored  cosmological model underpredicts the density of 
X-ray clusters with respect to the observations by a factor of $\sim 10$
for clusters with $T_x > 4 $ keV. The situation is
slightly better for the higher normalization   \MNUWH   simulation. But
still,  it predicts a factor of $\sim 6$  fewer density of clusters hotter
than $T_x>4$ keV than in reality.   

We showed in Fig \ref{massfunc} that effects of resolution are
negligible in the estimate of the cumulative mass function for 
  massive clusters. This could not be the case for the temperature
  estimates from the gas particles. In order to check whether the XTF
  could be affected by resolution, we also show in Fig \ref{xtf} a
  comparison of the XTF between \MNUC and
\MNUCL simulations.    As can be seen, the spectroscopic   
temperature estimate of clusters is biased high  when  low  mass
resolution is used  in a SPH simulation. 
Thus, we expect 
that  the difference  in  XTF shown  between \MNUC concordance model
simulation with $1024^3$ particles  and the   WMAP3 lower resolution
simulations ($512^3$)  is in fact a  lower limit.  
If we increased the mass resolution of the latter, we
  would  obtain a larger difference  with respect to the \MNUC and data.

\begin{figure}
\plotone{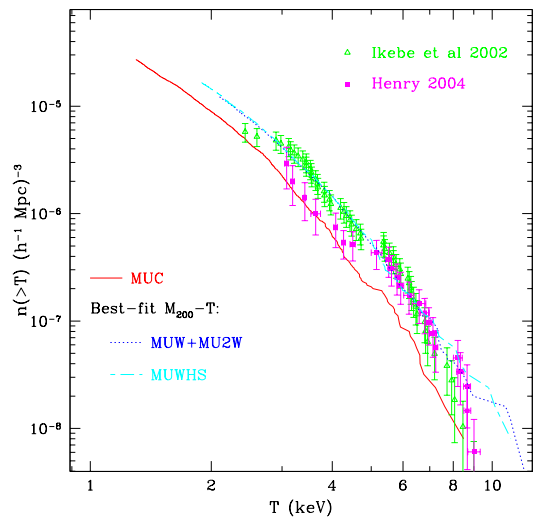}
\caption{ Best-fit cumulative  XTF  to data from 
WMAP3 simulation mass functions (see Fig \ref{massfunc})  and varying
the two free parameters of the $M_{200}-T_x$ relation. 
The  XTF   from the  \MNUC simulation  
is  shown for comparison.\label{xtffit} }
\end{figure}

\section{Discussion}

We have shown in the previous sections that the low normalized WMAP3
cosmological simulations underpredict the abundance of X-ray clusters
by a factor that ranges between $6-10$ with respect
 to estimates from ASCA  observations.  
Now,  our  estimates are based on the results from non-radiative 
gas dynamical  simulations  of the ICM.    There is still
no clear answer to what extent cooling and star formation  
 are important in the thermodynamics of the ICM.
 The extreme complexity of the processes involved  
presents a serious challenge for  simulating them accurately
in a  cosmological setting.  Results from simulations which
incorporate some modeling of these processes have shown that the M-T
relation is not strongly affected by non-gravitational heating 
\citep{borgani:04, nagai:07}.  A rather more important ingredient in the
determination of the XTF from mass functions is the intrinsic scatter 
of the M-T relation.  If the scatter is big, then a rather low
normalization power spectrum can in principle give high enough XTF to
be compatible with observations. Given the very good
statistical sample of objects in our simulations, we can reliably
estimate not only the M-T relation but also the intrinsic scatter due
to the  cluster dynamics.  In Table \ref{simu}, we report the 
least-square  fit values of the  $ M_{200}/M_\odot h^{-1} = (M_0/M_\odot h^{-1} ) (T_X/3
~\mbox{keV})^\alpha$ for the different simulations where  errors
in both the slope $\alpha$ and
normalization $M_0$ correspond  to $1 \sigma$ in the fit. 
We can also make a reliable estimate of the  intrinsic scatter in
the $M_{200}-T_X$ relation. The linear fit of the $\log M_{200} $ versus $\log
T_X$ has a Pearson's correlation coefficient  better than 0.99 for all
simulations. The maximum intrinsic scatter,  $\Delta \log M_0$, is also
shown in Table \ref{simu}. It is
defined as the value for which   99\% of all the clusters used in the
fit  have their spectroscopic temperature within
the values  $\log T_X = (\log M_{200}) /\alpha -\log M_0 \mp \Delta \log M_0 $.  As
can be seen, the  values of the scatter are between $0.28-0.31$ dex
($\sim  2 $ factor with respect to $M_0$) for the WMAP3 simulations. 
 But are the  differences shown in Fig \ref{xtf}   between  the simulated XTF
 and data  compatible with this intrinsic scatter of the M-T
 relation? In order to give a possible answer to this question,  we have
 estimated the  $M_o$ and $\alpha$ parameters of the $M_{200}-T_X$  relation needed 
to accommodate the  mass functions shown in Fig \ref{massfunc} to the
observational XTF data by a  $\chi ^2$ minimization. We show in Fig
\ref{xtffit} the best-fit simulated XTF for the WMAP3 simulations 
to the observational data points, together with the  simulation results for the
high-normalization  \MNUC simulation.  
The $\chi^2$ best fit  values found   for the   \MNUW + \MNGW
are  $\alpha^I =1.64$  and $\log M^I _0 =14.09 $ for the Ikebe et al data 
   and $\alpha^h = 1.49$ and $\log M^h _0 =14.17$  for the Henry data. When
   both observational data sets  are  taken together in the fit, we obtain
   $\alpha^{I+h}=1.64$,  $\log M^{I+h} _0 =14.10$.  For the
   higher normalization WMAP3 simulation  \MNUWH,  we find  $\alpha^I =1.66$;
   $\log M^I _0 =14.18 $ for Ikebe,   $\alpha^h = 1.44$ and $\log M^h _0
   =14.28$  for Henry and $\alpha^{I+h}=1.67$,  $\log M^{I+h} _0 =14.18$ for
   the combined datasets. Now, if  we fix the  slope, $\alpha$,   to the  best
 fit value  obtained from each simulation (see Table \ref{simu})
 we find  a value for   normalization parameter  $\log M _0
 =14.10-14.13$ for  WMAP3 $\sigma_8=0.75$  simulations and $\log M _0
 =14.19-14.22$ for the    $\sigma_8=0.8$ WMAP3 simulation.  Finally, 
if we assume the self-similar behavior  of the M-T scaling relation,
 $\alpha= 3/2$, then, the best  fit values
for $M_0 $ are  quite similar:   $\log M_0 =14.20-14.26$ for the \MNUWH
simulation and  $\log M_0 =14.12-14.17$ for the \MNUW + \MNGW simulations.
Therefore, the   normalization of the $M_{200}-T_x$  relation needed to fit the
observational XTF for  the $\sigma_8=0.75$   \MNUW and  \MNGW simulations is
a factor of $0.40-0.45$  dex  ($\sim 2.5-2.8 $ times ) smaller  than the best
fit values shown in Table \ref{simu}. For the $\sigma_8=0.8$  WMAP3 \MNUWH simulation this
factor is $0.36-0.39$ dex ($\sim2.3-2.4$ factor). As we have seen, the maximum
scatter derived from our WMAP3   non-radiative gas dynamical simulations is
$\sim \pm 0.28-0.31$ dex (i.e. a factor of $\sim 2$ ). 
   It is not clear  that non-gravitational heating 
could affect the thermodynamics of the ICM in
such a way that this   could account for a factor of $\sim 2.5-2.8$ lower  
 normalization  with respect to the predictions 
of the simulations reported here. For instance, the normalization for
the emission-weighted  $M_{500}-T_{ew}$  from SPH simulations including cooling and star
formation \citet{borgani:04} is a factor of 1.46 smaller than the value
we obtained for our  \MNUWH simulation.  
If we compare the normalization of the spectroscopic $M_{500}-T_x$ from
the  radiative cluster simulations of \citet{nagai:07} with ours, the
difference is within a factor of  $1.5-1.6$.

In conclusion, it seems unlikely  that we can reproduce the
observational estimates of abundance of X-ray clusters   with a
normalization of the power spectrum as low as the best fit value given
by WMAP3. A slightly higher normalization of $\sigma_8=0.8$ alleviates
the problem, although  the cluster abundance still lies below the observational
estimates. Considerably steeper slopes and lower  normalization of the
M-T relation are needed  to reconcile the predicted mass functions of
clusters with the observed XTF in this case.   Alternative explanations
which retain a low normalisation of $\sigma_8$  appeal to the effects of
 primordial non-gaussianity \citep{sadeh} or to
dynamical dark energy \citep{bartelmann}. However for the standard 
cosmological model,  X-ray clusters of
galaxies seem to  prefer  a higher  $\sigma_8$ than predicted by the CMB
 anisotropies,  in agreement  with the abundance of
optical clusters  from SDSS \citep{rozo}.

\vspace*{0.1cm}


The {\sc MareNostrum Universe} simulations have
 been done at BSC-CNS (Spain)   and  analyzed at NIC J\"ulich (Germany).
We thank A.I. Hispano-Alemanas and DFG for financial 
support. GY acknowledge support from    M.E.C. grants  
FPA2006-01105 and AYA2006-15492-C03. 
We thank P. Fosalba, A. Vikhlinin and W. Hu  for providing us 
 their  data.

\end{document}